\documentclass[fp,twocolumn]{jpsj3}
\usepackage{txfonts}
\usepackage[usenames, dvipsnames]{color}
\usepackage{epstopdf}
\newcommand{\beq}{\begin{eqnarray}}
\newcommand{\eeq}{\end{eqnarray}}

\title{First-principles Study of Spiral Spin Density Waves in Monolayer MnCl$_2$ Using Generalized Bloch Theorem}

\author{Teguh Budi Prayitno$^{1}$\thanks{teguh-budi@unj.ac.id}, Fumiyuki Ishii$^2$}
\inst{$^1$Department of Physics, Faculty of Mathematics and Natural Science, Universitas Negeri Jakarta, Kampus A Jl. Rawamangun Muka, Jakarta Timur 13220, Indonesia\\
$^2$Nanomaterials Research Institute, Kanazawa University, Kanazawa 920-1192, Japan} 

\abst{We investigated the spiral spin density waves in the monolayer 1T-MnCl$_2$ for a set of spiral vectors based on first-principles calculations. The magnetic ground states were evaluated by means of the generalized Bloch theorem within the linear combination of pseudo-atomic orbitals. To reach our purpose, a flat spiral configuration was constructed for the Mn magnetic atom by fixing the direction of its magnetic moment. We confirmed that the ground state was a spiral ground state. We also clarified that a phase transition from a spiral ground state to the other ground states, such as the ferromagnetic state or the antiferromagnetic state, appears when introducing the hole-electron doping. Therefore, we justify that introducing the hole-electron doping tunes the phase transition in the monolayer 1T-MnCl$_2$.}

\begin{document}
\maketitle

\section{Introduction}
\indent The investigation on the magnetic properties of the transition metal dihalides $XY_{2}$, where $X$ and $Y$ denote the metal cation and the halogen anion, is of interest due to various magnetic states, a good review can be found in Ref. \citen{1Mcguire}. Regarding these materials, both various ground states and multiferroic behaviors are verified by the experimental results \cite{2Kurumaji, 3Tokunaga, 4Kurumaji}. One of the great attention is addressed to the multiferroic behavior since it can be applied to the memory devices \cite{5Spaldin, 6Scott} by incorporating the ferroelectricity and the magnetism. This property suggests that the transition metal dihalides can be considered as one of the prominent candidates for the spintronics applications.       
    
Our concern is to consider one of the transition metal dihalides, i.e., MnCl$_2$, which can be crystallized in the CdCl$_2$ type structure with the appropriate space group $R\bar{3}m$. 
The important reason why we choose this material is due to the rare exploration of the magnetic order by using the density functional theory (DFT). The ground states of its bulk structure are reported to be two antiferromagnetic (AFM) transitions with the different low N$\acute{\textrm{e}}$el temperatures. Those two AFM states can be analyzed by the neutron scattering, however, the verification of these states needs a large cell within the DFT \cite{7Wiesler}. As a consequence, it is very difficult to construct the Heisenberg Hamiltonian to study the interaction of spins of atoms. On the other hand, the helimagnetic (HM) state has also been observed in the bulk structure. Based on the spin configurations, the HM order can be closely related to the spin spiral configuration \cite{Kurz}. Interestingly, some authors reported that the spiral configuration in the bulk system of MnI$_2$ can generate the ferroelectric polarization \cite{Wu}.

Beside the bulk structure, the magnetic order of the monolayer MnCl$_2$ is also a great interest. In contrast to the bulk structure, the monolayer MnCl$_2$ is reported experimentally from the bulk structure to have either the stripe order or the HM order \cite{1Mcguire}. It seems that the magnetic order of the monolayer MnCl$_2$ was deduced by analyzing the magnetic orders of the bulk structure. It is interesting because the Mermin$\textendash$Wagner theorem previously prohibited the magnetic order in the two-dimensional system \cite{Mermin}. Later, it was confirmed that the magnetic order in the two-dimensional system can also be induced by the magnetic anisotropy. Even though the monolayer MnCl$_2$ only considers the single layer, the calculation based on the DFT also requires a large cell to confirm these magnetic orders. We expect the new interesting phenomena can be observed due to the magnetic order in the monolayer MnCl$_2$. 

The main intention of this paper is to investigate the spiral (SP) ground state in the monolayer MnCl$_2$. Later, we also examine if the phase transition occurs when the doping is introduced. Previously, some authors reported that the phase transition can occur when increasing the hole doping \cite{8Inoue, 9Sawada}. Since in the monolayer MnCl$_2$ there is only one Mn$^{2+}$ cation, for the similar case, introducing the doping can be done experimentally, such as by a sol-gel method \cite{Du} or a hydrothermal method \cite{Zhang}. The main problem to investigate the SP state is to use a large cell, similar to the HM and the stripe orders. To reduce the computational cost, we applied the generalized Bloch theorem (GBT) using the primitive unit cell containing one Mn magnetic atom and two Cl nonmagnetic atoms. Due to its limitation regarding the orientation of the magnetic moment governed by the spiral vectors, we only considered the three stable states, i.e., the ferromagnetic (FM), SP, and AFM states. We observed that the SP ground state does exist for the nondoped case, while the other stable states can be tuned by introducing the doping. Therefore, we claimed that our calculation succeeds to prove the experimental result of the HM state in the monolayer MnCl$_2$.

We organize the rest of the paper as follows. The computational method and the crystal structure of the monolayer MnCl$_2$ will be discussed in Sec. 2. We also provide a detailed explanation of how to produce the FM, SP, and AFM states by setting the spiral vectors. The stability of the SP ground state will be qualitatively discussed by comparing with the previous study. In Sec. 3, the phase transition, which includes the three stable states, will be given in terms of the doping interval for the four different lattice constants. Then, the mechanism of the phase transition will be given by using the Heisenberg model in Sec. 4. We close our discussion by summarizing our results in Sec. 5.

\section{Method}
\indent We used the OPENMX code \cite{10Openmx}, a package for exploring material properties based on the DFT with the linear combination of pseudo-atomic orbitals (LCPAO) \cite{11Ozaki, 12Ozaki} as basis sets and the norm-conserving pseudopotentials \cite{13Troullier}, to investigate the spiral spin density waves (SSDW) in the monolayer MnCl$_2$. In an LCPAO, the SSDW in the materials can be expressed by the rotation of the magnetic moment of the magnetic atoms as
\beq
	\textit{\textbf{M}}_{i}(\textit{\textbf{r}}+\textit{\textbf{R}}_{i})=M_{i}(\textit{\textbf{r}}) \left(
\begin{array}{cc}
\cos\left(\varphi_{0}+\textit{\textbf{q}}\cdot \textit{\textbf{R}}_{i}\right)\sin\theta_{i}\\
\sin\left(\varphi_{0}+\textit{\textbf{q}}\cdot \textit{\textbf{R}}_{i}\right)\sin\theta_{i}\\
\cos\theta_{i}\end{array} 
\right), \label{moment}  
\eeq
where $\textit{\textbf{q}}$ and $\textit{\textbf{R}}_{i}$ are the spiral vector and the lattice vector at site $i$. Here, $\theta$ is the cone angle, which must be specified from the very beginning, between the spin rotation axis and the magnetic moment. By using the expression above, the Bloch wavefunction can be written in terms of a linear combination of pseudo-atomic orbitals (PAOs) $\phi_{i\alpha}$ at site $\tau_{i}$ as \cite{Teguh}
\beq
\psi_{\nu\textit{\textbf{k}}}\left(\textit{\textbf{r}}\right)&=&\frac{1}{\sqrt{N}}\sum_{n}^{N}\sum_{i\alpha}\left[e^{i\left(\textit{\textbf{k}}-\frac{\textit{\textbf{q}}}{2}\right)\cdot\textit{\textbf{R}}_{n}}C_{\nu\textit{\textbf{k}},i\alpha}^{\uparrow}
\left(
\begin{array}{cc}
1\\
0\end{array}
\right)\right.\nonumber\\
& &\left.+e^{i\left(\textit{\textbf{k}}+\frac{\textit{\textbf{q}}}{2}\right)\cdot\textit{\textbf{R}}_{n}}C_{\nu\textit{\textbf{k}},i\alpha}^{\downarrow}\left(
\begin{array}{cc}
0\\
1\end{array}
\right)\right]\nonumber\\
& &\times\phi_{i\alpha}\left(\mathrm{\textit{\textbf{r}}-\tau_{i}-\textit{\textbf{R}}_{n}}\right).\label{lcpao}
\eeq    
The complete explanation of implementing the GBT in the OPENMX code can be found in Ref. \citen{14Teguh}.

\begin{figure}[h!]
\vspace{1mm}
\centering
\includegraphics[scale=0.55, width =!, height =!]{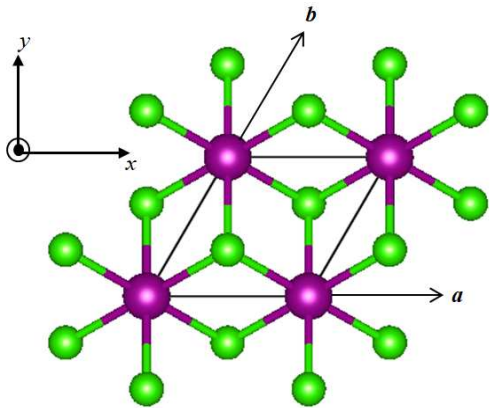}
\vspace{3 mm}
\caption{\label{cell}(Color online) Crystal structure of monolayer MnCl$_{2}$ from the top view. The black parallelogram denotes the unit cell, while the purple and green balls represent Mn and Cl atoms, respectively.} 
\end{figure}

\indent The monolayer MnCl$_2$ crystal with the space group $R\bar{3}m$ contains one Mn atom and two Cl atoms, as shown in Fig. \ref{cell}. To start the discussion, we set the experimental lattice constant of 3.686 {\AA} from the bulk structure and the length of vacuum ($z$ direction) of 17.47 {\AA} \cite{15Wyckoff}, see also Ref. \citen{16Tornero} for the comparison.          
Following Fig. \ref{cell}, we defined the primitive lattice vectors as
\beq
\textit{\textbf{a}}=a\hat{e}_{x}, \qquad  \textit{\textbf{b}}=\frac{a}{2}\hat{e}_{x}+\frac{a}{2}\sqrt{3}\hat{e}_{y},
\eeq 
where $a$ is the lattice constant. Then, the appropriate primitive reciprocal lattice vectors are found to be
\beq
\textit{\textbf{A}}=\frac{2\pi}{a}\hat{e}_{x}-\frac{2\pi}{a\sqrt{3}}\hat{e}_{y}, \qquad  \textit{\textbf{B}}=\frac{4\pi}{a\sqrt{3}}\hat{e}_{y}.
\eeq  

All the self-consistent calculations were performed using a $20 \times 20 \times 1$ \emph{k} point mesh in a primitive unit cell with the cutoff energy of 200 Ryd and the electronic temperature of 300 K. The functional of exchange-correlation was set to the generalized gradient approximation (GGA)\cite{17Perdew}. The basis set of Mn atom is specified by Mn4.0-$s$3$p$3$d$3$f$2, which means that three valence orbitals ($s$, $p$, and $d$ orbitals) and two polarization orbitals ($f$ orbital) were used, while the cutoff radius was set to 4.0 Bohr. Meanwhile, the basis set of Cl atoms was denoted by Cl7.0-$s$2$p$2$d$1, meaning that two valence orbitals ($s$ and $p$ orbitals) and one polarization orbital ($d$ orbital) were used, while the cutoff radius was set to 7.0 Bohr. 

\begin{figure}[h!]
\vspace{1mm}
\centering
\includegraphics[scale=0.55, width =!, height =!]{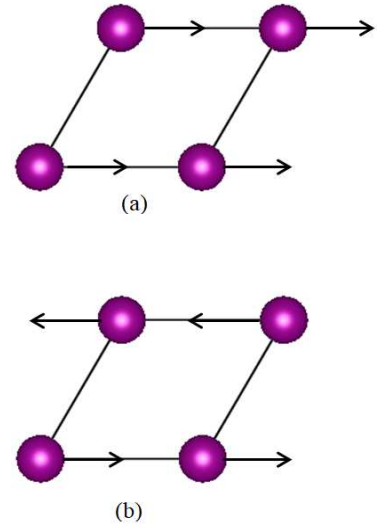}
\vspace{3 mm}
\caption{\label{FM-AFM}(Color online) Spin configurations of FM state with $\phi=0$ (a) and AFM state with $\phi=1$ (b).} 
\end{figure}

\indent To observe the SP ground state, a flat spiral configuration ($\theta=90^{\circ}$) for the Mn atom is arranged with the defined spiral vector $\textit{\textbf{q}}=\phi(\textit{\textbf{A}}+0.5\textit{\textbf{B}})$, where $\phi$ runs from 0 (FM state) to 1 (AFM state), as illustrated in Fig. \ref{FM-AFM}.
From Fig. \ref{FM-AFM}, the SSDW at various $q$ will then be determined in the interval between $\phi=0$ and $\phi=1$.  
             
\section{Results}
\subsection{Nondoped Case}
\begin{figure}[h!]
\vspace{-5mm}
\includegraphics[scale=0.55, width =!, height =!]{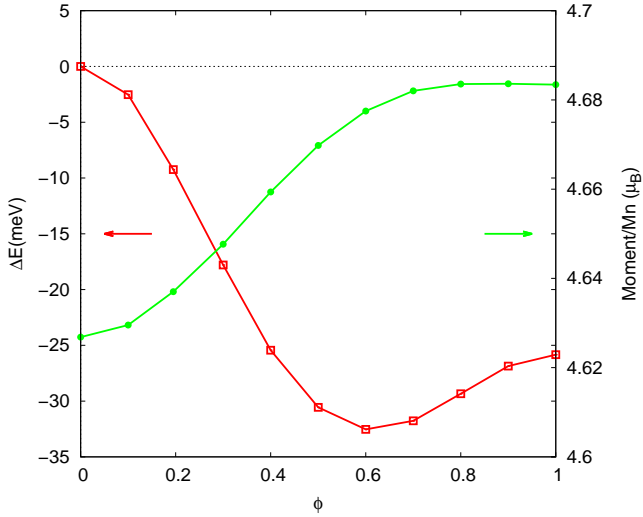}
\vspace{2 mm}
\caption{\label{spiral}(Color online) Total energy difference with respect to the FM state ($\phi=0$), $\Delta E=E(\phi)-E(\phi=0)$, and its corresponding magnetic moment of the SSDW using the lattice constant of 3.686 {\AA}. Empty squares and filled circles refer to the total energy difference and the magnetic moment.}
\vspace{-5 mm} 
\end{figure}

\indent We plot the total energy difference and the appropriate magnetic moment as the function of $\phi$, as shown in Fig. \ref{spiral}. As immediately observed in Fig. \ref{spiral}, the SP ground state occurs at $\phi=0.6$ with the magnetic moment of about 4.67 $\mu_{\textrm{B}}$. If the reciprocal lattice vectors are transformed to the Cartesian coordinates, the position of the SP ground state is quite same compared with that of the $\gamma$-Fe (fcc phase of iron), which also has an SP ground state \cite{18Uhl, 19Mryasov, 20Korling, 21Bylander1, Sandratskii, 22Bylander2, 23Sjo, 24Garcia}. Therefore, we use the $\gamma$-Fe as a reference to investigate the stability of the SP ground state in the monolayer MnCl$_{2}$. For this purpose, we introduce two kinds of the FM states, i.e., the low-spin (LS-FM) and high-spin (HS-FM) ferromagnetic states. Now, we confirm that the state with the magnetic moment larger than 4 $\mu_{\textrm{B}}$ is an HS-FM state.              

\begin{figure}[h!]
\vspace{-5mm}
\includegraphics[scale=0.55, width =!, height =!]{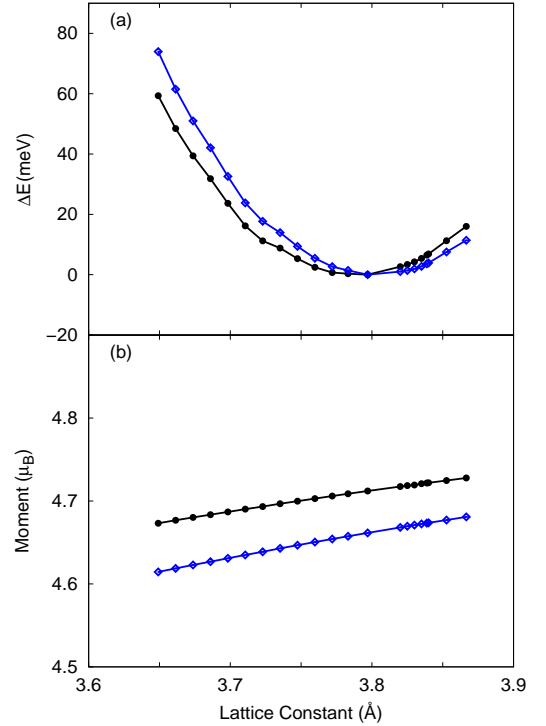}
\vspace{-14 mm}
\caption{\label{lattice_spin}(Color online) Lattice constant dependence of the total energy difference (a) and the magnetic moment (b) of the FM (diamonds) and AFM (filled circles) states. In this case, the total energy difference, $\Delta E_{FM(AFM)}=E_{FM(AFM)}(a)-E_{FM(AFM)}(a=3.797)$, is evaluated with respect to the minimum energy at the lattice constant of 3.797 {\AA}.}
\vspace{-2 mm} 
\end{figure}

\indent The existence of the SSDW in the $\gamma$-Fe is considered due to a crossing point between the HS-FM state and the AFM state. This crossing point can be regarded as a consequence of the stabilization of the $\gamma$-Fe at the low temperature. Another consequence is addressed to the sensitivity of the ground state of the $\gamma$-Fe to the lattice constant. Following this fact, our first attempt to investigate the stability of the SSDW in the monolayer MnCl$_{2}$ is to find a crossing point between the AFM state and, either the LS-FM state or the HS-FM state. To realize it, we graph the dependence of the total energy difference and the appropriate magnetic moment on the lattice constant for the FM and AFM states, as shown in Fig. \ref{lattice_spin}. In Fig. \ref{lattice_spin}, the atomic positions for all the lattice constants were optimized until the force acting on the atom is less than 0.05 meV/{\AA}. 

\begin{figure}[h!]
\vspace{-4mm}
\includegraphics[scale=0.55, width =!, height =!]{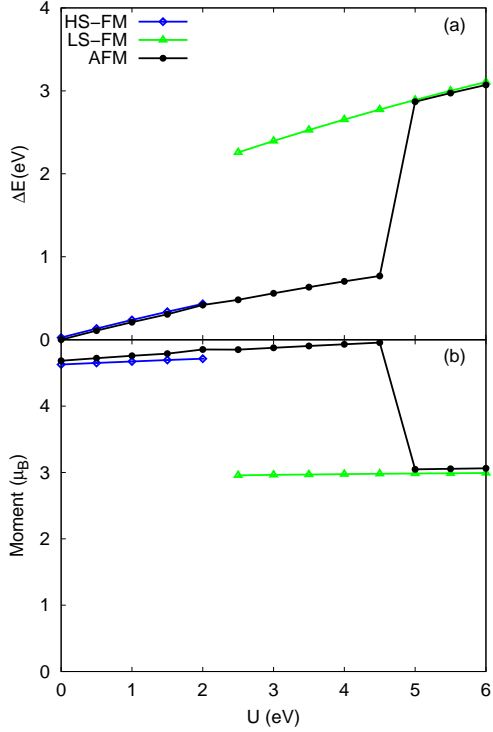}
\vspace{-14 mm}
\caption{\label{U}(Color online) $U$-dependent total energy difference (a), $\Delta E=E(U)-E(U=0)$, and magnetic moment for several states (b) using the lattice constant of 3.686 {\AA}.}
\vspace{-8 mm} 
\end{figure}

From Fig. \ref{lattice_spin}, we find a crossing point between the HS-FM and AFM states without observing the LS-FM state. This means that the SP ground state is sensitive to the lattice constant, similar to the $\gamma$-Fe. Furthermore, it can be seen that the AFM state becomes more stable than the HS-FM state when the lattice constant is less than 3.797 {\AA}. To convince our claim, we check and find that the FM ground state appears for the lattice constant larger than 4.2 {\AA}. Note that the sensitivity of the ground state to the lattice constant may possibly bring the sensitivity to the strain. Moreover, we also deduce that the optimized lattice constant is found to be 3.804 {\AA} by fitting the data of the dependence of the total energy on the lattice constant by using the collinear FM state. The LS-FM state can only appear, in our calculation, by applying the effective Coulomb energy $U$ in the implementation of the LDA+$U$ method in the OPENMX code \cite{25HanL1}. Figure \ref{U} shows the existence of the LS-FM state when applying $U>$ 2 eV. However, we cannot obtain, in this case, a crossing point between the FM state and the AFM state.  

Here, we would like to comment on the reliable $U$ value, which was used in the previous study, i.e., Mn$^{2+}$ system. By using the OPENMX code, Han et al. \cite{25HanL1} showed that the reliable value of $U$ lies between 4 eV and 6 eV to obtain the experimental gap of MnO system. Comparing to their result, we claim that the LS-FM state in this interval can be accepted, as shown in Fig. \ref{U}. Furthermore, since the AFM state is more stable than the FM state for the lattice constant of 3.686 {\AA}, the magnetic moment of the FM state reduces faster than that of the AFM state, where their transitions have the different critical $U$ value, see Fig. \ref{U}(b).          
     
Note that if one wants to discuss the exchange interaction in the monolayer MnCl$_{2}$, it seems to follow the so-called Goodenough-Kanamori-Anderson (GKA) rules \cite{26Goodenough, 27Kanamori, 28Anderson}. The original GKA rules explore the two different kinds of the superexchange interactions, i.e., the FM and AFM superexchange interactions, based on the angle of magnetic ion-ligand-magnetic ion. These two magnetic ions are referred to the partially filled $d$ orbitals, such as Mn ion. An FM superexchange interaction works if the angle is $90^{\circ}$ while an AFM superexchange interaction takes place if the angle is $180^{\circ}$. However, it was reported that a kind of materials, such as CuGeO$_{3}$ \cite{29Hase, 30Hori, 31Castilla}, sometimes violates the original rules, see ref. \citen{32Geertsma}. For the case of monolayer MnCl$_{2}$, a kind of superexchange interactions can be analyzed by considering the angle of Mn-Cl-Mn, which is about $101^{\circ}$.         

\subsection{Doped Case}
\begin{figure}[h!]    
\vspace{2mm}
\includegraphics[scale=0.6, width =!, height =!]{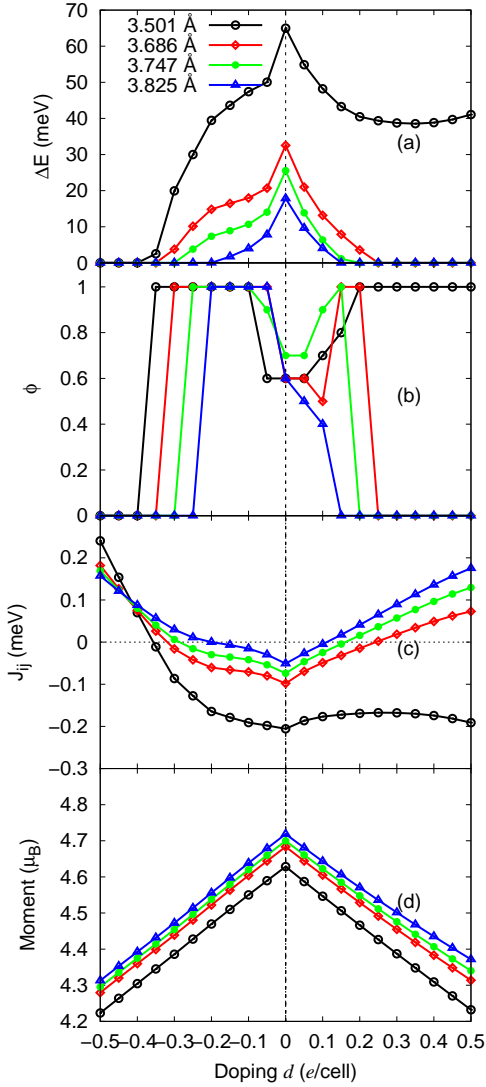}
\vspace{-16 mm}
\caption{\label{phase}(Color online) Phase transition in the monolayer MnCl$_{2}$ in the doping interval $d$ ($e$/cell) for the four lattice constants, as shown in (a) and (b). For each doping, the total energy difference (a), $\Delta E=E(a,d,\phi=0)-E(a,d,\phi=\phi_{\textrm{lowest}})$, is evaluated by subtracting the energy of the lowest state $E(\phi=\phi_{\textrm{lowest}})$ from the energy of FM state $E(\phi=0)$ for each $a$ and $d$. Meanwhile, $\phi$ represents the lowest state for each doping (b). Tendencies of the exchange parameter and the magnetic moment, as shown in (c) and (d). The exchange parameter (c), $J_{ij}=(1/12) [E(a,d,\phi=1)-E(a,d,\phi=0)]/M^{2}$, and the magnetic moment at $\phi=1$ in the monolayer MnCl$_{2}$ (d), in the doping range $d$ ($e$/cell) for the four lattice constants. The same tendencies for the other $\phi$ are also confirmed.}
\vspace{-8 mm} 
\end{figure}

For the next discussion, we investigate the phase transition in the monolayer MnCl$_{2}$ by applying the hole-electron doping. Figure \ref{phase} shows the doping-dependent ground state for the monolayer MnCl$_{2}$. For simplicity, we express the concentration of the doping per cell as $d$ ($e$/cell). As shown in Fig. \ref{phase}(a), the total energy difference increases for all cases (nondoped and doped cases) as the lattice constant increases. To start the discussion on the phase transition, as shown in Fig. \ref{phase}(b), let's consider first the lattice constant of 3.686 {\AA}, as represented by the diamonds in Fig. \ref{phase}. $0\leq d\leq$ 0.1 $e$/cell is the interval, at which the ground state of the system is an SP state. As the hole doping increases from 0 to 0.5 $e$/cell, the AFM state appears in the range of 0.15 $e$/cell $ \leq d \leq$ 0.2 $e$/cell while the FM state becomes the ground state in the range of $d \geq$ $0.25$ $e$/cell. As the electron doping increases, the AFM state appears in the range of $-0.3$ $e$/cell $ \leq d \leq -0.05$ $e$/cell while the FM state becomes the ground state in the range of $d \leq$ $-0.35$ $e$/cell. 

Based on the explanation above, we expose the phase transition for the other lattice constants. We select the lattice constants of 3.501 {\AA}, 3.686 {\AA}, 3.747 {\AA}, and 3.825 {\AA} to investigate the tendency of the phase transition as well as the competition between the superexchange interaction and the double exchange interaction in the next discussion. First of all, the SP state occurs for the region close to the nondoped case for all the lattice constants. It is also shown that the FM state becomes almost stable for all the lattice constants for the electron doping less than $-0.3$ $e$/cell. The significant change occurs for the AFM state, which is very sensitive to the doping. As the lattice constant decreases at 3.501 {\AA}, as represented by the empty circles in Fig. \ref{phase}(b), the AFM state becomes dominant for the hole doping at $d\geq$ 0.2 $e$/cell and appears in the interval of $-0.35$ $e$/cell $ \leq d \leq -0.1$ $e$/cell for the electron doping. Furthermore, the FM state acquires the small portion for only the electron doping in the interval of $d \leq$ $-0.4$ $e$/cell. On the contrary, as the lattice constant increases at 3.747 {\AA}, represented by the filled circles in Fig. \ref{phase}(b), the AFM state acquires only the small portion for the hole doping at $d=0.15$ $e$/cell and appears in the interval of $-0.25$ $e$/cell $ \leq d \leq -0.1$ $e$/cell for the electron doping. In addition, the FM state acquires the large portion for the hole doping in the interval of $d\geq$ 0.2 $e$/cell and the electron doping in the interval of $d\leq$ $-0.3$ $e$/cell. The same tendency also occurs at 3.825 {\AA}, represented by the triangles in Fig. \ref{phase}(b), which is close to the optimized lattice constant. We observe that the AFM state appears in the interval of $-0.2$ $e$/cell $ \leq d \leq -0.05$ $e$/cell for the electron doping, but no AFM state appears for the hole doping. In addition, the FM state also gains the large portion in the interval of $d\geq$ 0.15 $e$/cell for the hole doping and $d\leq$ $-0.25$ $e$/cell for the electron doping.      

Based on the above results, we clarify that the transformation of the ground state of FM-AFM-SP-AFM-FM occurs when varying $d$ from $-0.5$ $e$/cell to 0.5 $e$/cell. The existence of the phase transition in the monolayer MnCl$_2$ on the doping can be simply understood by the Heisenberg model
\beq
E=E_{0}-\frac{1}{2N} \sum_{i\neq j}J_{ij}\textit{\textbf{M}}_{i}\cdot\textit{\textbf{M}}_{j}
		\label{Heisenberg} 
\eeq 
where $N$ corresponds to the number of unit cells. According to Ref. \citen{Torun}, in which the authors discussed the 1T monolayer FeCl$_{2}$, the exchange parameter is given by $J_{ij}=(1/12)\Delta E_{xc}/M^{2}$. Here, the multiplier 1/12 is intended to overcome the double counting in the summation because one Mn atom is surrounded by six nearest neighbour Mn atoms. Moreover, we choose $M$ as the magnetic moment of the AFM state and $\Delta E_{xc}$ as the total energy difference between the AFM and FM states since the AFM state is more stable than the FM state for the nondoped case. The trends of the exchange parameter $J_{ij}$ as well as the magnetic moment can be seen in Figs. \ref{phase}(c) and \ref{phase}(d). The positive $J_{ij}$ represents the FM state while the negative one denotes either the AFM state or the SP state, as shown in Fig. \ref{phase}(c). Meanwhile, Fig. \ref{phase}(d) shows that increasing the doping will reduce the magnetic moment.

\section{Discussions}
By following Ref. \citen{8Inoue}, we would like to give a qualitative explanation about two different interactions, which change the ground state. As previously mentioned in Sec. 3, a superexchange interaction controls the magnetic properties of MnCl$_{2}$ for the nondoped case. This interaction can be considered as an indirect interaction because the interaction is mediated by Cl$^{-}$ ion as a nonmagnetic ion, which is located between two Mn$^{2+}$ ions. In this case, each Mn$^{2+}$ ion contains three electrons in the $t_{2g}$ state and two electrons in the $e_{g}$ state, where all the electron spins should have the same direction to obey the Hund's rule. At the same time, the on-site Coulomb repulsion will prevent two electrons having the same direction in the $e_{g}$ state for the nearest neighbour Mn atoms. So, the magnetic moments for the nearest neighbour Mn atoms should be antiferromagnetically coupled. This delocalizes the electrons over Mn-Cl-Mn, thus allowing the electron hopping from an Mn atom to the nearest neighbour Mn atom. As a consequence, the kinetic energy reduces, a loss of the kinetic energy.  

For the doped case, the FM state will be created by the so-called double exchange interaction, in which the magnetic moments in the neighboring Mn atoms are ferromagnetically coupled. Consequently, this interaction will prohibit the electron hopping between the nearest neighbour Mn atoms, thus the kinetic energy remains unchanged, a gain of the kinetic energy. On the contrary, the AFM and SP states will be induced by the superexchange interaction. This means that a gain of the kinetic energy favors an FM order while a loss of the kinetic energy leads to either an AFM order or an SP order. The ground state will be then determined by the domination between the superexchange and double exchange interactions. If the superexchange interaction is more dominant than the double exchange interaction, it leads to either an AFM state or an SP state; otherwise it favors an FM state. 

By using the explanations above, the superexchange interaction is more dominant than the double exchange interaction in the interval of $d \geq$ $-0.35$ $e$/cell for the lattice constant of 3.501 {\AA}, $-0.3$ $e$/cell $\leq d \leq$ 0.2 $e$/cell for the lattice constant of 3.686 {\AA}, $-0.25$ $e$/cell $\leq d \leq$ 0.15 $e$/cell for the lattice constant of 3.747 {\AA}, and $-0.2$ $e$/cell $\leq d \leq$ 0.1 $e$/cell for the lattice constant of 3.825 {\AA}. Meanwhile, the double exchange interaction is more dominant than the superexchange interaction in the interval of $d \leq$ $-0.4$ $e$/cell for the lattice constant of 3.501 {\AA}, $d \leq$ $-0.35$ $e$/cell and $d \geq$ 0.25 $e$/cell for the lattice constant of 3.686 {\AA}, $d \leq$ $-0.3$ $e$/cell and $d \geq$ 0.2 $e$/cell for the lattice constant of 3.747 {\AA}, and $d \leq$ $-0.25$ $e$/cell and $d \geq$ 0.15 $e$/cell for the lattice constant of 3.825 {\AA}. We deduce that the domination of the double exchange interaction increases as the lattice constant increases. Therefore, it is consistent with the appearance of the FM ground state at the lattice constant larger than 4.2 {\AA}. 

When the distance of Mn-Mn increases as the lattice constant increases, the electron will be difficult to hop between the nearest neighbour Mn atoms, thus the kinetic energy almost remains unchanged. This difficulty is influenced by the hopping integral $t$, which determines how much energy is required for an electron to hop from one site to the other site. In this case, the strength of the superexchange interaction is proportional to $t^{2}$, whereas the strength of the double exchange interaction is only proportional to $t$. Consequently, as the lattice constant increases, the strength of the superexchange interaction reduces more rapidly than that of the double exchange interaction. Therefore, this is the reason why the double exchange interaction is more dominant than the superexchange interaction at the large lattice constant.

The competition between the superexchange and the double exchange interactions also translates the critical doping. As shown in Fig. \ref{phase}(c), the critical doping decreases as the lattice constant increases. At the same time, introducing the doping tends to decrease the superexchange interaction to enter the double exchange region. Therefore, the decrease of the superexchange interaction leads to the decrease of the critical doping as the lattice constant increases. Note that the trend of $J_{ij}$ does not change for the hole doping case at the lattice constant of 3.501 {\AA}. This lattice constant is too small compared with the optimized lattice constant, thus the superexchange interaction is much more dominant than the double exchange interaction. Furthermore, we cannot also observe a hole-electron symmetry, as shown in Fig. \ref{phase}(c). This may be caused by the different orbital occupation of the hole and electron doping. When the hole doping is taken into account, it occupies the $e_{g}$ state. Contrarily, the $t_{2g}$ state will be occupied by the electron when the electron doping is introduced. Therefore, the trend of $J_{ij}$ will be different for the hole and electron doping. 
     
\section{Conclusions}
\indent We verify the SSDW in the monolayer MnCl$_{2}$, as predicted in the experimental result using the bulk structure, by using the GBT. For the nondoped case, the SP ground state is sensitive to the lattice constant due to a crossing point between the HS-FM state and the AFM state. We also show that the LS-FM state can only be attained by increasing the effective Coulomb energy $U$, however, no crossing point can be observed. In this case, we justify that the stability of the SP state depends on the lattice constant.                       

\indent By introducing the doping, the phase transition appears from the SP-AFM-FM states in general although we cannot see the FM state at the lattice constant of 3.501 {\AA} for the hole doping. These states can be tuned in the range of doping, as shown in Figs. \ref{phase}(a) and \ref{phase}(b). We also justify that the appearance of the phase transition in the monolayer MnCl$_{2}$ is due to the competition between the superexchange and double exchange interactions.   
  
\section*{Acknowledgment}
The computational calculations were partly carried out using ISSP supercomputers at the University of Tokyo while the remaining calculations were conducted using the server computer at the Universitas Negeri Jakarta. This work was supported by Japan Society for the Promotion of Science (JSPS) Grants-in-Aid for Scientific Research on Innovative Area, 'Nano Spin Conversion Science' (Grant Nos. 15H01015 and 17H05180). It was also supported by a JSPS Grant-in-Aid for Scientific Research on Innovative Area, 'Discrete Geometric Analysis for Material Design' (Grant No. 18H04481). It was partially supported by a JSPS Grant-in-Aid on Scientific Research (Grant No. 16K04875)

\end{document}